\documentclass{iopconfser}
\usepackage{amsmath}
\usepackage{graphicx}
\usepackage{orcidlink}

\DeclareMathOperator{\cosech}{cosech}
\usepackage{xcolor}

\begin{document}

\title{Quantum-corrected three-dimensional AdS space-time}

\author{Jacob C. Thompson\ \orcidlink{0000-0003-1047-9257} and Elizabeth Winstanley\ \orcidlink{0000-0001-8964-8142}}

\affil{School of Mathematical and Physical Sciences, University of Sheffield, Hicks Building, Hounsfield Road, Sheffield S3 7RH, United Kingdom.}

\email{JThompson16@sheffield.ac.uk, E.Winstanley@sheffield.ac.uk}

\begin{abstract}
We study quantum-corrected solitons in global, three-dimensional, anti-de Sitter (AdS) space-time.
These static solitons have a regular origin and arise as solutions of the linearized quantum-corrected Einstein equations (LQCEE). 
On the right-hand-side of the LQCEE is the renormalized expectation value of the stress-energy tensor operator for a massless, conformally coupled, quantum scalar field in a nonrotating thermal state,  
computed in quantum field theory (QFT), or using relativistic kinetic theory (RKT). 
We calculate the mass of the solitons and compare the results from QFT and RKT. 
\end{abstract}

\section{Introduction}
\label{sec:intro}
In general relativity, all energy and matter gravitates, and thus results in the curvature of space-time, governed by the classical Einstein equations:
\begin{equation}
    G_{\mu \nu }+ \Lambda g_{\mu \nu } = T_{\mu \nu },
    \label{eq:EE}
\end{equation}
where $G_{\mu \nu }$ is the Einstein tensor for the metric $g_{\mu \nu }$, the energy and matter distribution is described by the stress-energy tensor $T_{\mu \nu }$, and $\Lambda $ is the cosmological constant (we use units in which $8\pi G = c=\hbar = k_{{\mathrm {B}}}=1$).
Let us consider a simple type of matter field, namely a massless, conformally-coupled scalar field.  
In four space-time dimensions, the only asymptotically flat solutions of the Einstein equations (\ref{eq:EE}) with $T_{\mu \nu }$ for such a scalar field are black holes \cite{Xanthopoulos}.
However, in asymptotically anti-de Sitter (AdS) space-time with $\Lambda <0$ such a scalar field can
also generate soliton-like solutions \cite{Radu:2005bp}.
In three dimensions, regular vortex solutions with a charged classical scalar field can also be constructed in AdS \cite{Edery:2020kof,Edery:2022crs}.

A natural question is whether solitons can also be generated when the classical scalar field is replaced by a massless, conformally-coupled, \emph{quantum} scalar field. 
In this  situation the classical stress-energy tensor $T_{\mu \nu }$ in (\ref{eq:EE}) is replaced by the renormalized expectation value of the stress-energy tensor operator (RSET) $\langle {\hat {T}}_{\mu \nu }\rangle $.
The RSET is computed on a \emph{fixed} AdS background, and a suitable quantum state has to be chosen for the scalar field.
On pure AdS space-time, the RSET for the global vacuum state results simply in a renormalization of the cosmological constant $\Lambda $ \cite{Kent:2014nya}, so we consider instead a global thermal state.
In four dimensions, it is found \cite{Thompson:2024akz} that the thermal RSET $\langle {\hat {T}}_{\mu \nu } [g^{0}]\rangle $ (where $g^{0}$ is the background AdS metric) generates \emph{quantum-corrected}  (QC) solitons.  

In this note, we consider the simpler situation of three-dimensional AdS, inspired by the work of \cite{Casals:2019jfo} on quantum-corrections to the three-dimensional, asymptotically AdS, BTZ black hole. 
While in \cite{Casals:2019jfo} the background space-time possesses a singularity,
here we work on the regular global AdS space-time.
In our previous work \cite{Thompson:2024akz}, we solved the nonlinear Einstein equations. 
However, it is expected that the RSET (and therefore quantum corrections to the AdS metric) will be small, of order the Planck length $\ell _{\mathrm {P}}$.
We therefore follow \cite{Casals:2019jfo} and take a perturbative approach, writing the QC metric as $g^{\mathrm {QC}} = g^{0} + \ell_{\mathrm {P}} \, \delta g + {\mathcal {O}}(\ell_{\mathrm {P}}^{2})$, and solve the ${\mathcal {O}}(\ell_{\mathrm {P}})$ linearized quantum-corrected Einstein equations (LQCEE)
\begin{equation}
    \delta G_{\mu \nu } + \Lambda \,\delta g _{\mu \nu } = \langle {\hat {T}}_{\mu \nu } [g^{0}]\rangle ,
    \label{eq:LQCEE}
\end{equation}
where $\delta G_{\mu \nu }$ is the linearized perturbation of the Einstein tensor. 
In the following we briefly review the properties of the RSET source $\langle {\hat {T}}_{\mu \nu } [g^{0}]\rangle $ \cite{Thompson:2025jkn} before discussing soliton-like solutions of the LQCEE (\ref{eq:LQCEE}). 
Here we consider only static solitons, and we plan to return to rotating solitons in a future work \cite{Thompson}.

\section{RSET on AdS${}_{3}$}
\label{sec:rset}
\begin{figure}[h]
    \centering
    \includegraphics[width=0.327\linewidth]{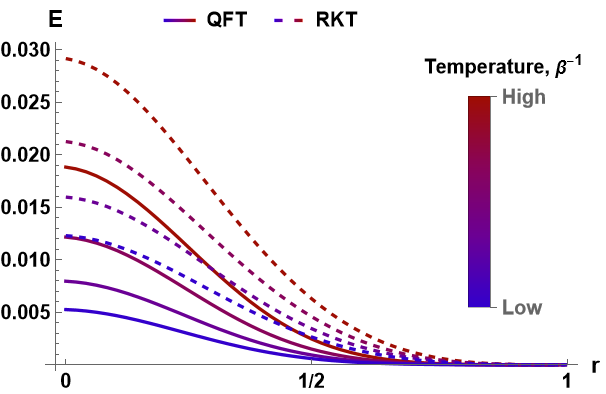}
    \includegraphics[width=0.327\linewidth]{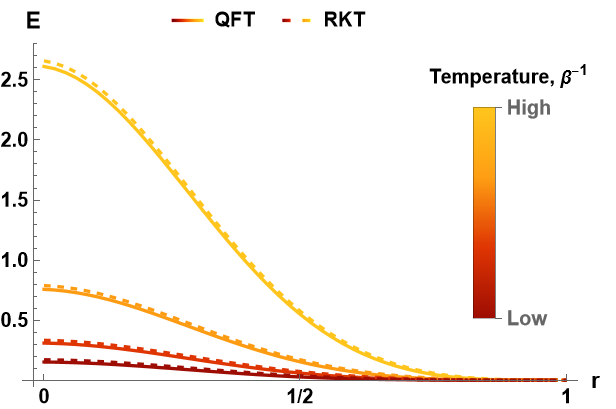}
    \includegraphics[width=0.327\linewidth]{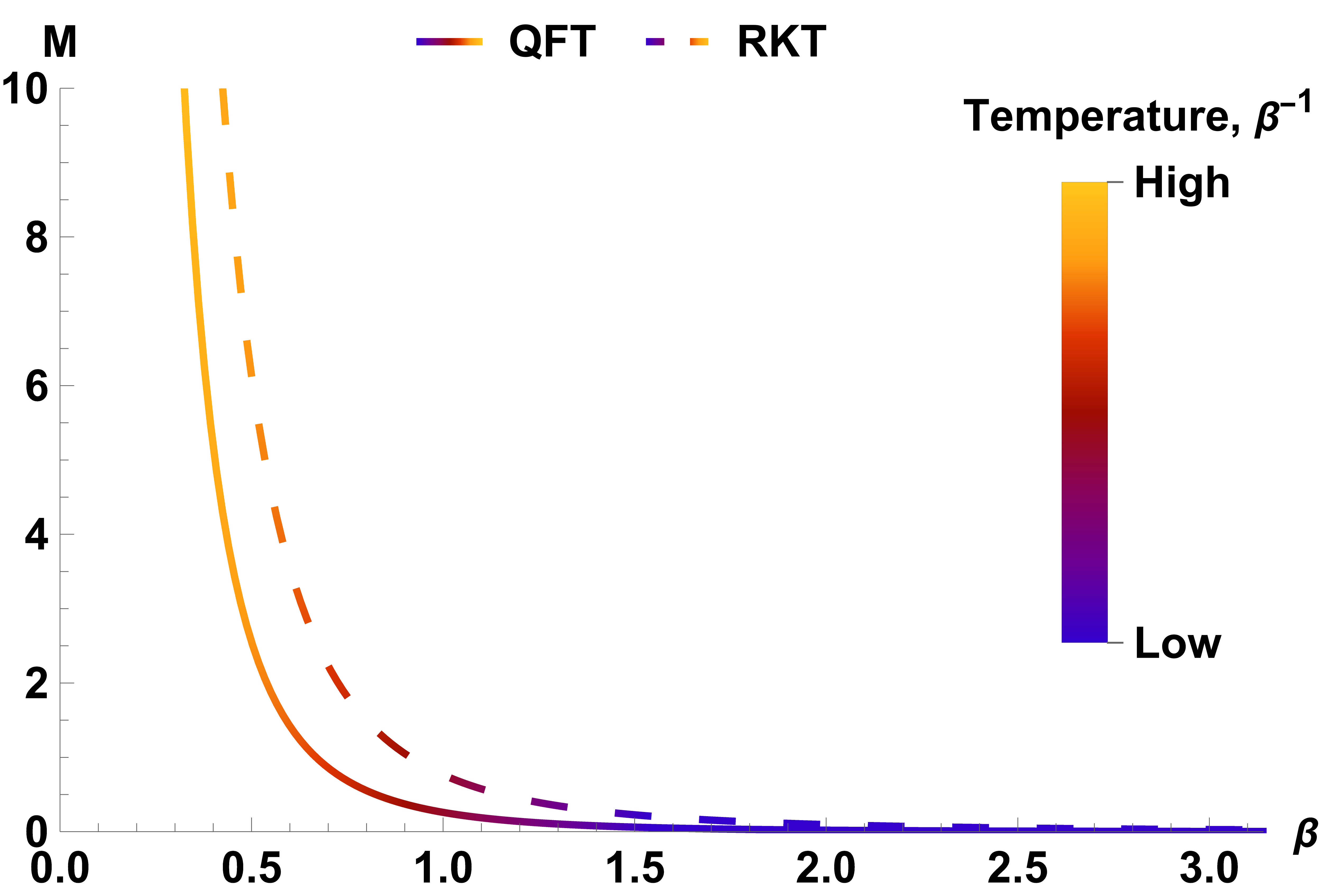}
    \caption{Energy density $E=-T_{t}^{t}$ for $\beta \in \{3\pi/4, 5\pi/6, 11\pi/12, \pi \}$ (left) and $\beta \in \{ \pi/6, \pi /4, \pi /3, 5\pi /12 \}$ (middle). 
    QC soliton mass $M$  (\ref{eq:mass}) (right) as a function of inverse temperature $\beta $. 
    Dotted lines denote results from RKT, while solid lines are those from QFT. We set the AdS inverse length scale $a=1$.}
    \label{fig:plots}
\end{figure}

Our background is three-dimensional global AdS space-time with metric $g^{0}_{\mu \nu }$ given by the line element
\begin{equation}
       {\mathrm {d}}s^{2}_{0} = a^{-2}\left( 1 - r^{2} \right) ^{-2} \left[ 
   - \left( 1 + r^{2}\right) ^{2} {\mathrm {d}}t^{2} 
+4 \, {\mathrm {d}}r^{2} + 4r^{2} \,{\mathrm {d}}\theta^{2} 
   \right] ,
   \label{eq:metric}
\end{equation}
where $a$ is the inverse AdS length scale, related to the negative cosmological constant $\Lambda $ 
by $a={\sqrt {-\Lambda }}$ and the coordinates have ranges $-\infty < t < \infty $, $0\le r < 1$ and $0\le \theta < 2\pi $.
We consider a massless, conformally-coupled, quantum scalar field in a thermal state on the background metric (\ref{eq:metric}). 
The quantum field theory (QFT) expectation value $\langle {\hat {T}}_{\mu \nu } [g^{0}]\rangle$ is computed in \cite{Thompson:2025jkn} for a rigidly-rotating thermal state at inverse temperature $\beta $ (using a different coordinate system to that employed here). 
Setting the angular speed of rotation $\Omega $ to zero, and changing to the coordinate system in (\ref{eq:metric}), the QFT-RSET has nonvanishing components ${}^{\mathrm {QFT}}T_{t}^{t}$, ${}^{\mathrm {QFT}}T_{r}^{r}$ and ${}^{\mathrm {QFT}}T_{\theta }^{\theta }$ given by
\begin{subequations}
\label{eq:QFT}
\begin{align}
    {}^{\mathrm {QFT}}T_{r}^{r}  & = \frac{a^{3}}{16{\sqrt {2}}\pi }\frac{\left( 1 - r^{2}\right)^{3}}{\left( 1 +r^{2}\right)^{3}}
    \sum _{j=1}^{\infty }\left[ \frac{1}{4{\sqrt{2}}}\left[ 3 + \cosh \left( j\beta \right)\right] \cosech ^{3} \left( \frac{j\beta }{2}\right)
    - {\mathcal {A}}_{j}^{-\frac{3}{2}}\sinh ^{2}\left( \frac{j\beta }{2}\right) \right] ,
\\
    {}^{\mathrm {QFT}}T_{\theta }^{\theta } & = 
     {}^{\mathrm {QFT}}T_{r}^{r}-
    \frac{3a^{3}r^{2}}{2{\sqrt {2}}\pi }\frac{\left( 1 - r^{2}\right)^{3}}{\left( 1 +r^{2}\right)^{5}} \sum _{j=1}^{\infty }{\mathcal {A}}_{j}^{-\frac{5}{2}}\sinh ^{2}\left( \frac{j\beta }{2}\right) ,
\end{align}
\end{subequations}
with ${}^{\mathrm {QFT}}T_{t}^{t} + {}^{\mathrm {QFT}}T_{r}^{r}+{}^{\mathrm {QFT}}T_{\theta }^{\theta } =0 $ and ${\mathcal {A}}_{j}=\cosh\left(j\beta \right) + \left( 1 +r^{2}\right)^{-2}\left(1 - 6r^{2} + r^{4} \right)$.
Since the expressions (\ref{eq:QFT}) are somewhat unwieldy, we also take an alternative approach, namely relativistic kinetic theory (RKT). 
In RKT, the quantum scalar field is modelled as a classical gas of massless bosonic particles with a Bose-Einstein distribution function at inverse temperature $\beta $.
The resulting classical RKT-SET takes the much simpler form  (here $\zeta (3)$ is the value of the Riemann-zeta function) \cite{Thompson:2025jkn}
\begin{equation}
    {}^{\mathrm {RKT}}T_{\mu }^{\nu } = \frac{a^{3}\zeta (3)}{2\pi \beta ^{3}} \frac{\left( 1 - r^{2}\right)^{3}}{\left( 1 + r^{2}\right)^{3}}{\mathrm{Diag}} \left\{ -2, 1, 1 \right\} .
    \label{eq:RKT}
\end{equation}
In Fig.~\ref{fig:plots} we show the energy density $E=-T_{t}^{t}$ for low temperatures (left-hand-plot) and high temperatures (middle plot), with the RKT results given by dashed lines and the QFT results by solid lines.
At high temperatures (small $\beta $), the RKT-SET (\ref{eq:RKT}) is a very good approximation to the QFT-RSET (\ref{eq:QFT}), but it ceases to be a good approximation at low temperatures (large $\beta $) \cite{Thompson:2025jkn}.
For all fixed temperatures, the energy density computed using RKT is larger than in QFT at the same value of the radial coordinate $r$.
In both QFT and RKT, the energy density is positive throughout the space-time and has a soliton-like profile: for fixed inverse temperature $\beta $, the energy density has a maximum at the origin $r=0$ and decreases to zero as $r\rightarrow 1$ and the AdS boundary is approached \cite{Thompson:2025jkn}.

\section{QC AdS${}_{3}$ solitons}
\label{sec:solitons}

We now use the QFT-RSET (\ref{eq:QFT}) and the RKT-SET (\ref{eq:RKT}) as source terms in the LQCEE (\ref{eq:LQCEE}) and seek soliton-like solutions of these equations. 
We assume that the QC metric is static and spherically-symmetric, and takes the form
\begin{equation}
    {\mathrm {d}}s^{2}_{\mathrm {QC}} =a^{-2} \left[ -A(r) \, {\mathrm {d}}t^{2} + B(r) \, {\mathrm {d}}r^{2} + C(r) \, {\mathrm {d}} \theta ^{2}\right] .
    \label{eq:QCmetric}
\end{equation}
We write each of the metric functions $A(r)$, $B(r)$ and $C(r)$ as their expressions for the background metric (\ref{eq:metric}) plus small ${\mathcal {O}}\left( a\ell _{\mathrm {P}}\right)$ perturbations $\delta A(r)$, $\delta B(r)$ and $\delta C(r)$.
By making a suitable perturbation of the radial coordinate $r$, we may take the perturbation of the metric function $B(r)$ to be
\begin{equation}
    \delta B(r) = \frac{\left(1-r^{2}\right)^{2}}{r^{2}\left(1+r^{2}\right)^{2}}\delta A(r) .
\end{equation}
With this choice, the LQCEE (\ref{eq:LQCEE}) reduce to two ordinary differential equations, a second order equation for $\delta A(r)$ and a first order equation for $\delta A(r)+\delta C(r)$.
Employing, respectively, the method of variation of parameters and an integrating factor, these can be solved to give explicit expressions for the metric perturbations in terms of integrals over the RSET components (for details, see \cite{Thompson}):
\begin{subequations}
\label{eq:LQCsol}
\begin{align}
    \delta A(r) & = {\mathcal{C}}_{1} + \frac{{\mathcal{C}}_2 r^{2}}{\left(1-r^{2}\right)^{2}} 
    - \int \frac{1+r^{2}}{ar\left(1-r^{2}\right)}T_{\theta\theta}\, {\mathrm {d}}r 
    + \frac{r^{2}}{\left(1-r^{2}\right)^{2}} \int \frac{\left(1-r^{4}\right)}{ar^{3}}T_{\theta\theta} \, {\mathrm {d}}r ,
\\
    \delta A(r) + \delta C(r) & = \frac{r^{2}}{\left(1-r^{2}\right)^{2}}\left[ {\mathcal{C}}_{3} + 2\int\frac{1-r^{4}}{ar}T_{rr}\, {\mathrm {d}}r \right] ,
\end{align}
\end{subequations}
where ${\mathcal {C}}_{1}$, ${\mathcal {C}}_{2}$ and ${\mathcal {C}}_{3}$ are arbitrary constants, which are fixed by the requirement that the metric perturbations are regular at the origin $r=0$ and that the QC space-time approaches pure AdS as $r\rightarrow 1$.
Substituting the RKT-SET (\ref{eq:RKT}) and QFT-RSET (\ref{eq:QFT}) into the integrals in (\ref{eq:LQCsol}), explicit algebraic expressions for the metric perturbations can be obtained \cite{Thompson}, which are quite complicated in the QFT case. 

While the metric ansatz (\ref{eq:QCmetric}) is useful for explicitly finding the metric perturbations, in order to aid our soliton intepretation of the QC metric, we change to a new radial coordinate $R$, given by  $R^{2} = C(r)\in [0,\infty )$
and consider the alternative metric (following the QC BTZ ansatz in \cite{Casals:2019jfo}):
\begin{equation}
     {\mathrm {d}}s^{2}_{\mathrm {QC}} =a^{-2} \left[ 
     -N(R) f(R) \, {\mathrm {d}}t^{2} + N(R)^{-1} \, {\mathrm {d}}R^{2} + R^{2} \, {\mathrm {d}}\theta ^{2}
     \right],
    \label{eq:BTZ}
\end{equation}
where $N(R) = R^{2}+1+a\ell _{\mathrm {P}}m(R)+{\mathcal {O}}\left( a^{2}\ell _{\mathrm {P}}^{2}\right) $ and $f(R)=1+{\mathcal {O}}\left( a\ell _{\mathrm {P}}\right) $.
It is the metric perturbation $m(R)$ which is of particular interest for our purposes. 
This is given as a linear combination of the metric perturbations $\delta A$ and $\delta C$ and satisfies the differential equation $m'(R)=2RE/a^{2}$, where $E=-T_{t}^{t}$ is the local scalar field energy density \cite{Thompson}.
In the BTZ black hole metric, $m(R)$ is a constant, corresponding to  the mass of the black hole \cite{Banados:1992wn}. 
In order to have a regular origin at $R=0$ (which corresponds to $r=0$), it must be the case that $m(0)=0$
(which is not satisfied for the BTZ black hole, hence there is a singularity at the origin). 
The condition $m(0)=0$ is satisfied by an appropriate choice of the integration constants in (\ref{eq:LQCsol}) \cite{Thompson}.
Using both the RKT-SET and QFT-RSET in the LQCEE, since $E>0$, we find that $m(R)$ is a monotonically increasing function of $R$ and tends to a finite limit as $R\rightarrow \infty $ (which corresponds to the AdS boundary at $r=1$) due to the rapid decrease of $E$ as $R\rightarrow \infty $.  
By analogy with the BTZ metric \cite{Banados:1992wn}, we interpret this limit as the mass $M$ of the QC solitons, which takes the simple expressions
\begin{equation}
M = \lim _{R\rightarrow \infty } m(R) = \begin{cases}
     \,  {\displaystyle {\dfrac{2\zeta (3)}{\pi \beta ^{3}}}}, &  {\mathrm {RKT}},
     \\
    \,   {\displaystyle {\dfrac{1}{32\pi }\sum _{j=1}^{\infty } \left[ 3+ \cosh\left( j\beta \right) - \sinh \left( j\beta \right)\right] \cosech ^{3} \left(  \frac{j\beta }{2} \right) }},  & {\mathrm {QFT}},
\end{cases}    
\label{eq:mass}
\end{equation}
when evaluated using the RKT-SET and QFT-RSET, respectively.
The soliton mass $M$ is shown in the right-hand plot in Fig.~\ref{fig:plots} as a function of the inverse temperature $\beta $.
It can be seen that the mass computed using the RKT-SET is larger than that derived from the QFT-RSET, as expected since the local energy density $E$ is greater in RKT than QFT. 
Furthermore, the mass increases rapidly as the temperature of the thermal state increases.

\section{Conclusions}
\label{sec:conc}

In this note we have studied the backreaction of a massless, conformally-coupled, quantum scalar field in a static thermal state on three-dimensional global AdS space-time. 
Using a perturbative approach, we have solved the linearized, quantum-corrected, Einstein equations (LQCEE) to find the metric perturbations of pure AdS, with the scalar field RSET acting as a source. 
We have computed the RSET in QFT and compared this with an SET derived using RKT.
We find globally regular, static soliton solutions of the LQCEE, whose mass increases rapidly as the temperature of the thermal state increases.  
We are currently working on the corresponding rotating quantum-corrected solitons sourced by a rigidly-rotating thermal state on three-dimensional AdS, and plan to report our results in the near future \cite{Thompson}.

\section*{Acknowledgements}
We thank Siva Namasivayam for helpful discussions.
The work of J.T.~is supported by an EPSRC studentship.
The work of E.W.~is supported by STFC grant number ST/X000621/1.


\begin{thebibliography}{99}

\bibitem{Xanthopoulos}
Xanthopoulos B C and Zannias T  1991 {\it {J. Math. Phys.}} \textbf {32} 1875--1880 

\bibitem{Radu:2005bp}
Radu E and Winstanley E 2005
{\it {Phys. Rev. D}} \textbf{72} 024017 

\bibitem{Edery:2020kof}
Edery A 2021
{\it {JHEP}} \textbf{01} 166 

\bibitem{Edery:2022crs}
Edery A 2022
{\it {Phys. Rev. D}} \textbf{106} 065017 

\bibitem{Kent:2014nya}
Kent C and Winstanley E 2015
{\it {Phys. Rev. D}} \textbf{91} 044044

\bibitem{Thompson:2024akz}
Thompson  J C and Winstanley E 2024
{\it {Phys. Rev. D}} \textbf{110} 125003

\bibitem{Casals:2019jfo}
Casals M, Fabbri A, Mart{\'\i}nez C and Zanelli J 2019
{\it {Phys. Rev. D}} \textbf{99} 104023

\bibitem{Thompson:2025jkn}
Thompson J C and Winstanley E 2025
{\it {Phys. Lett. B}} \textbf{868} 139780

\bibitem{Thompson}
Thompson J C and Winstanley E,  work in progress.

\bibitem{Banados:1992wn}
Banados M, Teitelboim C and Zanelli J 1992
{\it {Phys. Rev. Lett.}} \textbf{69} 1849--1851

\end{thebibliography}
\end{document}